\documentclass[aip,jcp,amsmath,amssymb,a4paper,floatfix,longbibliography, reprint]{revtex4-1}

\usepackage{graphicx}
\usepackage{dcolumn}
\usepackage{bm}
\usepackage{subfigure}
\usepackage{xcolor}
\usepackage[english]{babel}
\usepackage[hidelinks]{hyperref}
\usepackage{mhchem}

\DeclareMathOperator{\sign}{sign}
\newcommand \textss \textsuperscript
\newcommand{\ud}{\mathrm{d}}

\newcommand{\figurewidth}{0.49\textwidth}

\begin{document}

\title[]{Jellium and Cell Model for Titratable Colloids with
  Continuous Size Distribution}

 \author{Guillaume Bareigts}
 \email{guillaume.bareigts@u-bourgogne.fr}
 \affiliation{ICB UMR 6303 CNRS, Univ. Bourgogne Franche-Comt\'e,
   FR-21000 Dijon, France}
 \author{Christophe Labbez}
 \email{christophe.labbez@u-bourgogne.fr}
 \affiliation{ICB UMR 6303 CNRS, Univ. Bourgogne Franche-Comt\'e,
   FR-21000 Dijon, France}

\date{\today}

\begin{abstract}
  A good understanding and determination of colloidal interactions is paramount to comprehend and model
  the thermodynamic and structural properties of colloidal
  suspensions. In concentrated aqueous suspensions of colloids with a
  titratable surface charge, this determination is, however, complicated by the density
  dependence of the effective pair potential due to both the many-body
  interactions and the charge regulation of the colloids. In addition,
  colloids generally present a size distribution which
  results in a virtually infinite combination of colloid pairs. In this paper we
  develop two methods and describe the corresponding algorithms to
  solve this problem for arbitrary size distributions. An
  implementation in Nim is also provided.
  The methods, inspired by the seminal work of Torres \textit{et al.}, are based
  on a generalization of the cell and renormalized jellium models to polydisperse suspensions of
  spherical colloids with a charge regulating boundary condition. The
  latter is described by the one-pK-Stern model. The predictions of
  the models are confronted to the equations of state of various commercially
  available silica dispersions. The renormalized Yukawa parameters
  (effective charges and screening lengths) are also calculated. The
  importance of size and charge polydispersity as well as the validity
  of these two models are discussed in light of the results.
\end{abstract}


\maketitle

\section{Introduction}

Size polydispersity, rather than being an exception, is a general rule
in the realm of colloidal systems. It has been shown to influence the
micro structure of suspensions \cite{DAguanno:91}, to considerably
enrich the number of crystal phases observed
\cite{Shevchenko:06,Evers:10,Cabane:16,Schaert:17}, to affect
nucleation\cite{Auer:01,Linden:13,Palberg:16}, to induce the fractionation
of particles during
crystallization\cite{Dijkstra:94,Bolhuis:96a,Kofke:99,Byelov:10,Cabane:16} and to
allow particular behavior such as colloidal Brazil nut effect
\cite{Esztermann:04,Noguera-Marin:15}, colloidal stratification
\cite{Fortini:16,Zhou:17} and fluid-fluid
demixing\cite{Aarts:05,Dijkstra:98,Zvyagolskaya:11}. Moreover,
polydispersity has been shown to be an essential feature in the
formation of colloidal glasses\cite{Pusey:86,Barrat:86} and has allowed substantial achievements
in the understanding of this state, see e.g. Refs \onlinecite{Fernandez:07,Pusey:09,Palberg:16a,Berthier:17a}.

Despite its ubiquity, polydispersity is still often neglected,  with the
exception of neutral hard sphere systems, and the variety of phases,
states and behaviors that it brings about are imperfectly controlled and
understood. The main reason for this is the fact that computer simulations
still lag well behind experimental observations, when appropriate
models exist at all. This is particularly true for charged
colloidal suspensions, the system of interest in this paper. 

From a simulation point of view, representative and realistic models
of charged polydisperse colloidal suspensions are indeed tractable neither
at the primitive model level of approximation, where the degrees of
freedom of the solvent molecules are averaged out through a dielectric
continuum, nor at the level of the mean field approximation\cite{Fushiki:92,Dobnikar:03,Hallez:14}, where the many
ions are further replaced by a mean electrostatic potential obtained
from solving the Poisson-Boltzmann equation. An amenable and attractive
approach, first introduced by Beresford-Smith \cite{Beresford-Smith:85a}, consists, instead,
in whittling the colloidal system down to the colloidal particles
only, i.e. the one-component model, while the degrees of freedom of the ions and solvent
molecules 
are integrated out in \textit{effective}
pair potentials between the colloids, $w^{\ast}(r)$. The
reduction of the many-body interactions into effective potentials,
however, makes $w^{\ast}$ density dependent \cite{Brunner:02} and, thus, necessarily
re-determined for each colloid density.

In the case of monodisperse spherical particles at low
electrostatic coupling, Alexander \textit{et al.} \cite{Alexander:84} showed
that $w^{\ast}(r)$ retains a simple Yukawa form but with
\textit{effective} parameters, namely
an effective charge, $Z^{\ast}$, and screening length, $1/\kappa^{\ast}$,
instead of the bare charge, $Z$ and bulk screening length, $1/\kappa$. The study further
showed that $Z^{\ast}$ and $\kappa^{\ast}$ can be obtained from solving the
Poisson-Boltzmann equation around \textit{one} colloid placed in Wigner-Seitz
cell model (CM)\cite{Marcus:55a}. In the same spirit, a one-colloid renormalized
jellium model was developed and shown to be successful in salt free
systems \cite{Trizac:04}. At high electrostatic coupling, \textit{two}-colloids cell \cite{Turesson:12}
and jellium models \cite{Bareigts:17} solved in the full primitive model were shown to provide accurate
$w^{\ast}(r)$ for arbitrary colloid shapes \cite{Thuresson:14} and concentrations. The two-colloid
approach was further used at a molecular level \cite{Bonnaud:16} in a 3D-periodic
simulation box.

In the case of charged colloid mixtures, Torres \textit{et al.} \cite{Torres:08} proposed a
generalization of the cell model. The great insight of Torres and
co-workers was simply to impose the same potential at the boundary of each cell, each family
of colloidal particles being represented by one colloidal
particle of the same radius centered in its own Wigner-Sietz cell, in such a way as to
ensure the continuity of electric potential and ion concentrations
across the cell boundaries. The greatest ideas also being the simplest, it
was then followed to generalize the jellium model \cite{Falcon-Gonzalez:11,GarciadeSoria:16}. However, to our
knowledge the generalized cell and jellium models have only
been tested in the salt free case \cite{GarciadeSoria:16}.
Furthermore, they have so far always been restricted to binary mixtures, i.e. have never
been applied to polydisperse charged colloidal suspensions with
continuous size distribution.

Another difficulty arises from the nature of the surface charge and its
dependence on the density and size polydispersity of colloidal suspensions, namely the \textit{charge regulation} and the
\textit{charge polydispersity}. Both largely depend on the chemistry of the colloid surface and
of the electrolyte but also on the surface curvature and strength of
the interactions and are, thus, specific to each colloidal system. The aqueous surface chemistry and charging behavior of
colloidal particles has been the subject of many investigations
essentially concerning the thermodynamic limit of infinite (colloid)
dilution, see e.g. Refs. \onlinecite{Hunter:95,Hiemstra:96,Borkovec:97c,Lyklema:05}. On the contrary, in studies of the structure of colloidal
suspensions, the charging behavior (of colloids) is most often simply ignored
the assumption being a constant surface density or at best a constant electrical double
layer potential \cite{Smallenburg:11}. This can be explained in part by the complexity
of characterization and, thus, by the poor knowledge of the charging
behavior of colloids in \textit{non diluted} suspensions, not to mention the
charge polydispersity, as indicated by the very limited research work available \cite{Wette:02,Lobaskin:07,Merrill:09}. Very
rarely have attempts been made to include a description of surface
chemistry \cite{Gisler:94,vonGrunberg:99,Heinen:14}. Furthermore, those that do exist
are, again, all limited to monodisperse systems.

Motivated by the recent experimental results obtained by Cabane
\textit{et al.}\cite{Cabane:16} on aqueous suspensions of titrating
silica nanoparticles with large polydispersity, which
show a fractionation of particles in three coexisting phases
(Laves/BCC/liquid phases) in the semi-concentrated regime and high pH, we here develop two methods and
describe the corresponding algorithms to estimate the charging
behavior and charge polydispersity of titrating silica particles with a continuous size distribution. The methods are further used to evaluate the sets of effective parameters
(i.e. $Z^{\ast}$ and $\kappa^{\ast}$) to be used in a one-component model. 
The methods, largely inspired by the seminal work of
Torres \textit{et al.}\cite{Torres:08}, are based on a generalization of the
cell and renormalized jellium models to polydisperse suspensions of
spherical colloids supplemented with a charge regulating boundary
condition described by a 1-p$K$-Stern model. Certain features are studied, in particular, the dependence of
the charge polydispersity as well as its scaling with the
surface curvature on the size polydispersity and density
of the colloids. Finally, the validity of the proposed models is discussed in terms of their ability to
describe the equation of state of various commercially available
silica dispersions.

The manuscript is organized as follows: in Sect. \ref{sec:models} we introduce the
models used, that is, in Sect. \ref{sec:pcm} and \ref{sec:prjm}, the generalized cell model and jellium
model for charged polydisperse colloidal suspensions and, in Sect. \ref{sec:BC}, the 1-pK Stern
model to describe the interface between the solid and the electrolyte
solution in the presence of acidic surface groups.
In Sect. \ref{sec:algorithm} the
algorithm used to solve the cell and jellium models coupled with the
1-pK Stern model is described.
In Sect. \ref{sec:charging_silica}  we present the 1-p$K$-Stern
model fit of the charging behavior of silica surfaces in the
dilute regime together with the CM and RJM predictions of the bare charge
polydispersity of silica nanoparticles with various polydispersities
 and densities, and in
 various pH conditions.
 The corresponding effective charge
polydispersities and effective screening lengths are presented in
Sect. \ref{sec:ren_params}. Microion pressures for various polydispersities and
distribution shapes is studied in Sect \ref{sec:osmotic}. Finally, experimental data are compared with the
predictions of the cell and jellium model in the same section, followed by conclusions in
Sect. \ref{sec:conclusion}.

\section{Models}
\label{sec:models}
Let us consider a polydisperse colloidal suspension composed of $n_p$
spherical colloidal species of radii $R_p$ bearing a charge $Q_p=Z_pe$ with
$e$ the elementary charge and $p=1,\dots,n_p$. They are immersed in a volume $V$ filled with an
aqueous salt solution of dielectric constant $\epsilon$ in equilibrium
with a reservoir at a temperature $T$ and of inverse screening length,
\begin{equation}
\label{kapres}
\kappa = \sqrt{4 \pi \lambda_B \sum_{i=1}^{n_i} z_i^2 c_{s,i}} ,
\end{equation}
where $\lambda_B=\frac{e^2}{4 \pi k_B T \epsilon}$ is the Bjerrum
length and $k_B$ is the
Boltzmann constant while $z_i$ and $c_{s,i}$ are the number valence and bulk
concentration of ionic species $i$, respectively. $n_i$ is the total number of ion species.
The composition of each colloidal species is defined by its number fraction
$x_p=N_p/\sum_{n_p}N_p$.

Within the mean-field approximation of the primitive model, the
electrostatic potential at the surface of the colloids, at a set
configuration of the latter, and in the
electrolyte solution is determined by solving the
Poisson-Boltzmann equation, which for an arbitrary system is given
by\cite{Israelachvili:91}
\begin{equation}
\label{PB}
\epsilon \bigtriangleup V(\vec{r}) + \sum_{i=1}^{n_i} z_ie c_i(\vec{r}) + \rho_e(\vec{r}) = 0 ,
\end{equation}
where $\vec{r}$ is the vector position in the solution, $V$ is the electrostatic potential,  
$\bigtriangleup$ is the Laplace operator, $c_i(\vec{r}) = c_{s,i} \exp \left ( \frac{ - z_ie V(\vec{r})}{k_B T} \right )$ and
$\rho_e$ is a charge density associated with the colloids, specified later according to the model.

Within the approximation of the polydisperse cell model (PCM) and
polydisperse renormalized jellium model (PRJM) it is only necessary to solve
Equation \ref{PB} with one colloid with the appropriate boundary
conditions and to repeat it for each colloid
species. Taking further advantage of the spherical symmetry, the
electrostatic potential becomes a mere function of the radial coordinate $r$ and
Eq. \ref{PB} reduces to 
\begin{equation}
\label{PBrad}
 \frac{\partial^2 \psi}{\partial r^2} + \frac{1}{r}\frac{\partial \psi}{\partial r}
+ 4 \pi \lambda_B \left [ \sum_{i=1}^{n_i} z_i c_{s,i} \exp( - z_i \psi(r) ) + \xi(r)  \right ] = 0 ,
\end{equation}
where for convenience, we have introduced the dimensionless potential
$\psi = \frac{e V}{k_B T}$ and the reduced charge
density $\xi = \rho_e/e$. The surrounding colloids are effectively accounted for through
the boundary conditions and $\xi$ which depends on the model
used. They are detailed below.

\subsection{Cell model}
\label{sec:pcm}

The cell model approximation emerged was initially designed for colloidal
crystals and emerged from the realization that due to its periodicity the
volume of a crystal can be divided into electroneutral Wigner-Seitz cells
surrounding each colloid which on average have the same volume and
contain the same ion concentrations\cite{Alexander:84}. In other words, the thermodynamic
properties of the system can be reduced to one colloid
enclosed in an appropriate cell. The geometry of the cell is further
assumed to have the same shape as the colloid. A spherical cell of
radius $R_c$ centered on the colloid is a natural choice for a
spherical colloid. Note that the cell model approximation was also shown to
be valid for non spherical colloids and moderately concentrated fluid
states\cite{Reiner:91,Lowen:94,Bocquet:02}.

Within this approximation, $\xi = 0$ and, thus, the PB equation,
Eq. \ref{PBrad}, within the electrolyte solution takes the usual form
\begin{equation}
  \bigtriangleup \psi(r)=\kappa^2\sinh\psi(r)\text{    with }R_p < r < R_c .
\end{equation}
Note that here a 1-1 salt solution is considered. The Gauss law imposes that the electric field be null everywhere on
the boundary of the electroneutral cell,
\begin{equation}
\label{bccm}
  \frac{\partial \psi}{\partial r} \Big |_{r=R_c} = 0 .
\end{equation}
The missing boundary condition at the colloid surface is described
below (Sect. \ref{sec:BC}).

For monodisperse dispersions the cell radius is commensurate
with the particle volume fraction $\Phi$,
\begin{equation}
  \label{phi}
  \Phi = \left ( \frac{R}{R_c} \right )^3 .
\end{equation}

Similarly, for polydisperse dispersions, the cell radii of the
colloidal species, $R_{c,p}$, are related to
the overall particle volume fraction by
\begin{equation}
\label{vfCM}
\Phi = \frac{\sum_{p=1}^{n_p} x_p R_p}{\sum_{p=1}^{n_p} x_p R_{c,p}^3} .
\end{equation}
These unknown variables are determined by imposing the continuity of
the electrical potential and ion concentrations over the
different cells\cite{Torres:08}. That is to say,
\begin{equation}
\label{closcm}
\psi(R_{c,1}) = \psi(R_{c,2}) = \dots = \psi(R_{c,n_p}) = \psi_c .
\end{equation}

In the case of suspensions of colloids immersed in mono-monovalent salt solutions, the effective pair potential between the
colloids keeps the form of a screened Coulomb potential,
\begin{equation}
 \label{yuk}
\beta u(r_{pq}) =\lambda_B \upsilon_pZ^{\ast}_{p}\upsilon_qZ^{\ast}_{q} \frac{\exp({-\kappa^{\ast} r})}{r},
\end{equation}
but with the renormalized charges and inverse screening length as
leading parameters. In the previous equation
$\upsilon_p=\exp(\kappa^{\ast}R_p)/(1+\kappa^{\ast}R_p)$ ensures that the
ionic cloud is excluded from the core of the colloid. Following the elegant method of Trizac \textit{et al.}\cite{Trizac:03}
those renormalized parameters can be obtained
analytically from the calculated electrostatic potential at the edge of the cell. That is,
\begin{equation}
  \kappa^{\ast} = \sqrt{4 \pi \lambda_B \sum_{i=1}^{n_i} z_i^2 c_{s,i} \exp( - z_i \psi_c)}
\end{equation}
and
\begin{align}
  \label{zeffcm}
  Z^{\ast}_{p} & = \frac{\gamma_0}{\kappa^{\ast} \lambda_B} 
   \left [ ((\kappa^{\ast})^2 R_{c,p} R_p - 1)
    \sinh \kappa^{\ast} (R_{c,p} - R_p) \right. \nonumber \\
    & \left. + (\kappa^{\ast})^2+ \kappa^{\ast} (R_{c,p} - R_p) \sinh  \kappa^{\ast} (R_{c,p} - R_p)
    \right ], 
\end{align}
where 
\begin{equation}
  \gamma_0 = \frac{- 4 \pi \lambda_B}{(\kappa^{\ast})^2}  \sum_{i=1}^{n_i} z_i c_{s,i} \exp( - z_i \psi_c),
\end{equation}
from which the effective charge density for colloid $p$ can be defined as
\begin{equation}
\sigma^{\ast}_{p} = \frac{Z^{\ast}_{p}}{4 \pi R_p^2}.
\end{equation}

The osmotic pressure of the colloidal dispersion can be approximated by the cell
model and is given by
\begin{equation}
P = k_B T ( c_{coll} + c_{ions, in} - c_{ions, res} ),
\end{equation}
with $c_{coll}$ the concentration of the colloids, $c_{ions, res}$ the
ion concentration of the reservoir, and $c_{ions, in}$ the ion
concentration of the dispersion, \textit{i.e.} the ion concentration at the edge
of the cell(s). The latter can be re-expressed as
\begin{equation}
\label{pcm}
P = k_B T \left \{ \frac{\Phi}{\bar{v}_p} + \sum_{i=1}^{n_i} c_{s,i} \left [ \exp( - z_i \psi_c) -1 \right ]  \right \},
\end{equation}
where $\bar{v}_p = \sum_{p=1}^{n_p} x_p v_p$. This
approximation for the osmotic pressure neglects, however, the contribution of
the colloids and is valid for low ionic strength and/or relatively
large particle volume fraction only. For a detailed discussion see Refs. 
\onlinecite{Belloni:00, Dobnikar:06, Hallez:14, Hallez:17}.

\subsection{Renormalized jellium model}
\label{sec:prjm}

In contrast to the cell model, the jellium model\cite{Beresford-Smith:85} is based on the fact that,
 for diluted suspensions,
the colloid-colloid radial distribution function can be approximated to
$g_{pp}=1$. That is, the colloids can be seen as uniformly
distributed throughout the suspension. The small ions are, on the
other hand, strongly correlated with the colloids. Once again, the
colloidal suspension can thus be reduced to a one-colloid subsystem
immersed in an infinite sea of salt solution supplemented by a uniform
background charge, namely $\xi = \xi_{back}$ in Eq. \ref{PBrad} which becomes
\begin{equation}
\label{PBradRJM}
 \frac{\partial^2 \psi}{\partial r^2} + \frac{1}{r}\frac{\partial \psi}{\partial r}
+ 4 \pi \lambda_B \left [ \sum_{i=1}^{n_i} z_i c_{s,i}\exp( - z_i \psi(r) ) + \xi_{back} \right ] = 0 .
\end{equation}

The background charge represents nothing but the other colloids
bearing a charge $Z_{back}$ smeared out in space. In the case of
a mono-disperse colloidal suspension of radius $R$, the particle volume fraction is
thus a simple function of $\xi_{back}$. That is
\begin{equation}
\label{bgcdef2}
\xi_{back} = Z_{back} \frac{3\Phi}{4\pi R^3} .
\end{equation}
As noted by Trizac \textit{et al.}\cite{Trizac:04} in most of the cases $Z_{back}$ is different
from the bare charge of the colloids which must be renormalized by
fitting the electrostatic potential tail obtained by means of
Eq.\ref{PBradRJM} with the known far field expression for $\psi(r)$,
see below.

In order to keep the system electroneutral a
Donnan potential is set at infinite separation from the
colloid, \textit{i.e.} $\psi(\infty)=\psi_D$, given by
\begin{equation}
\label{defphiD}
  \xi_{back} = - \sum_{i=1}^{n_i} z_i c_{s,i} \exp(- z_i \psi_D) .
\end{equation}
Furthermore, the condition of electroneutrality imposes,
\begin{equation}
\label{bcrjm}
  \frac{\partial \psi}{\partial r} \Big |_{r \to +\infty} = 0 .
\end{equation}

The generalization of the renormalized jellium model to polydisperse
colloidal suspensions is obtained  simply by positing that the background charge
is the charge density caused by a \textit{uniform mixture} of
colloidal species $p$ bearing a charge $Z_{back,p}$, so that
\begin{equation}
\label{bgcdef3}
\xi_{back} = \frac{\Phi}{\bar{v}_p} \sum_{p=1}^{n_p} x_p Z_{back,p} ,
\end{equation}
where $\bar{v}_p = \frac{4}{3}\pi\sum_{p=1}^{n_p} x_p R_p^3$, or, equivalently,
that the overall colloidal volume fraction is given by,
\begin{equation}
  \label{vfRJM}
  \Phi = \frac{\bar{v}_p \xi_{back}
  }{\sum_{p=1}^{n_p} x_p  Z_{back,p} }.
\end{equation}
In other words, the continuity of the electrostatic potential and ion
concentrations is ensured by imposing the same $\xi_{back}$ for all
colloidal species $p$.

The $Z_{back,p}$ values are obtained by an iterative procedure which
consists in equating them to their respective effective charges,
$Z^{\ast}_{p}$, obtained from a fit of the tail of the far-field electrostatic
potential profile by the expression of the linearized potential, $\tilde\psi_p(r)$,
\begin{equation}
\tilde\psi_p(r) = \psi_D + \lambda_B \frac{Z^{\ast}_{p} }{1 + \kappa^{\ast} R_p} \frac{\exp({-\kappa^{\ast}(r-R_p)})}{r},
\end{equation}
where
\begin{equation}
  \kappa^{\ast} =  \sqrt{4 \pi \lambda_B \sum_{i=1}^{n_i} z_i^2 c_{s,i} \exp( - z_i \psi_D)},
\end{equation}
which thus gives a new value of $\xi_{back}$ and $\psi_p(r)$,
until convergence of $Z^{\ast}_{p}$ for all colloidal species $p$,

\begin{equation}
\label{rjmclose}
Z_{back,p}(Z^{\ast}_{p}) = Z^{\ast}_{p}\quad \forall p \in \{1,\dots,n_p\}.
\end{equation}

Similarly to the cell model (Eq \ref{pcm}), the osmotic pressure of
the colloidal dispersion can be expressed as
\begin{equation}
\label{prjm}
P = k_B T \left \{ \frac{\Phi}{\bar{v}_p}+ \sum_{i=1}^{n_i} c_{s,i} \left [ \exp( - z_i \psi_D) -1 \right ]  \right \}.
\end{equation}
Again, this expression neglects the contribution of the colloid-colloid
correlations to the osmotic pressure.

\subsection{Boundary conditions at the colloidal surface}
\label{sec:BC}

So far, we have introduced the equation governing the electrostatic potential in the solution and the boundary
conditions specific to the model used. In the following, we describe the
boundary conditions relative to the surface of the colloids.

\subsubsection{General conditions}
For a colloid dressed with a bare charge density $\sigma$ a
general boundary condition can be expressed from the Gauss theorem
\begin{equation}
\label{bcch}
  \frac{\partial \psi}{\partial r} \Big |_{r=R_p} = - 4 \pi \lambda_B \sigma,
\end{equation}
In the case of a titrating surface charge
 a more convenient condition can be obtained from the electroneutrality
condition and reads
\begin{equation}
\label{bcch2CM}
\sigma = \frac{1}{R_p^2}\int_{R_p}^{R_c}\ud r r^2\sum_{i=1}^{n_i} z_i c_{s,i} \exp( - z_i \psi(r) )
\end{equation}
for the cell model and
\begin{equation}
\label{bcch2RJM}
\sigma = \frac{1}{R_p^2} \int_{R_p}^{\infty} \ud r r^2 \left [ \sum_{i=1}^{n_i} z_i c_{s,i} \exp( - z_i \psi(r) ) - \xi_{back}
\right ]
\end{equation}
for the renormalized jellium model. The above boundary conditions,
although necessary to solve the cell model and the renormalized
jellium model do not prejudge (define) either the nature of the colloidal charge or the
response of the latter to colloid density or to a change in
the nature and concentration of the electrolyte solution.

\subsubsection{Titrating surface charge}\label{titradesc}
\label{sec:titration}

In the case of a chemically inert colloid surface two conditions can be
defined, namely the constant charge and constant potential
conditions\cite{Hunter:95}. The first condition, however, gives rise to a
nonphysical result when two such charged surfaces are in contact: the
osmotic pressure becomes infinite! On the contrary, as two colloids
approach, the second condition implies that $\sigma$ drops and
eventually completely vanishes on contact. The constant potential forms the
lower bond of the charge regulation condition. It can also be seen as
a cheap and implicit manner to qualitatively account for the chemistry
of the interface, namely here the binding of the counter-ions.

In reality, the chemistry of the solid/solution interface is specific to the nature
of both the surface and the electrolyte. This chemistry can be
specified/defined along the chemical reaction equilibrium with associated
Gibbs free energies. They are then coupled with the physical
interactions undergone by the reaction products and reactants to form the
generically-termed physical chemistry of interfaces. The
chemical reactivity, in some sense, gives a fourth dimension to, and thus
considerably enlarges the domain of possible states of colloidal
systems.

Let us consider the situation in which the colloids bear titratable
surface sites with a surface density $d_s$. The surface sites are
either in a protonated state, $\ce{M-OH^{q_s+}}$ with charge $q_s^{+}$,
or deprotonated state, $\ce{M-O^{q_s-}}$ with charge $q_s^{-}$,
depending on the equilibrium pH of the reservoir. Their partition can be
conveniently quantified by the ionization fraction, $\alpha=N_{\ce{M-O^{q_s-}}}/(N_{\ce{M-OH^{q_s+}}}+N_{\ce{M-O^{q_s-}}})$. The bare surface
density is then obtained from $\sigma=d_s(\alpha q_s^{-}+(1-\alpha)
q_s^{+})$. The change in charge state of the surface sites with pH
obeys the following chemical equilibrium
\begin{equation}
\label{equil}
\ce{M-OH^{q_s+} \ce{<=>} M-O^{q_s-} + H+},
\end{equation}
and associated equilibrium constant, a function of the Gibbs free
energy,
\begin{equation}
  \label{const}
K_a=\exp(-\beta\Delta G)=\frac{a_{\ce{M-O^{q_s^{-}}}}a_{\ce{H+}}}{a_{\ce{M-OH^{q_s^{+}}}}},
\end{equation}
where the $a_s$ represent the chemical activities. Using
the surface concentration for the definition of the standard
composition\cite{Kallay:04}, $\Gamma_s=d_s/N_A$ where $N_A$ is the
Avogadro number, the chemical activity of any chemical
species at the colloid interface can be written as
\begin{equation}
  \label{acti}
a_s=\Gamma_s\exp(q_s \psi_s),
\end{equation}
The fraction of deprotonated sites is obtained by combining
Eqs. \ref{const}-\ref{acti} and reads
\begin{equation}
\label{pHregul}
\ln\frac{\alpha}{1-\alpha} = \ln 10 \left
( \text{pH} - \text{p}K_a \right ) - \psi_s \left ( q_s^{-}-q_s^{+} \right ).
\end{equation}

Finally, a Stern layer of thickness $\lambda_{Stern}$ is introduced around each colloids to account
for the hydrated size of the ions and the hydration layer of the
colloids\cite{Brown:15}. The surface sites are considered to reside within this
layer, that is, on the unhydrated surface of radius
$R_p-\lambda_{Stern}$. Disregarding dielectric discontinuities,
$\psi_s$ can be deduced from the diffuse layer electric potential $\psi_d(R_p)$. It can be defined by the following expression
\begin{equation}
  \label{stern}
\psi_s = \psi(R_p) +  \frac{4 \pi \lambda_B \lambda_{Stern}}{1 + \lambda_{Stern}/ R_p} \sigma,
\end{equation}
obtained from the definition of the capacitance \cite{walker:14} for a
spherical particle. Eqs. \ref{equil}, \ref{pHregul}, \ref{stern} form the basis of
the 1-p$K$ Stern model.
With the model specific boundary condition (Eq. \ref{bccm} for the cell model
and Eq. \ref{bcrjm} for the renormalized jellium model),
Eq. \ref{PBrad} can be solved for each particle size at any given pH. 
The detailed algorithms are described in the next section.

\subsection{Algorithm description}
\label{sec:algorithm}

\paragraph{The Poisson-Boltzmann equation}\label{PBE}, thereafter referred
to PBE, was numerically solved with an ``in house'' code 
based on the Newton Gauss-Seidel method\cite{NumRec}, see Supp-Info for more details. 
For a particle of radius $R_p$, and for a given pH, the potential profile $\psi(r)$ is calculated by the following algorithm:
\begin{enumerate}
\item Choose a first guess for the potential at $r=R_p$, $\psi_d$, within a range [$\psi_d^m$, $\psi_d^M$] .
\item Solve the PBE with a given $\psi(R)=\psi_d$.
\item Compute $\sigma$ (Eq. \ref{bcch}-\ref{bcch2CM}) and  $\text{pH}(\psi_d)$ (Eq. \ref{pHregul}).
\item If $\text{pH}(\psi_d)-\text{pH}$ is small enough, stop the
  program and return the results.
\item Dichotomy step:

  if $\sign[\text{pH}(\psi_d) - \text{pH}] = \sign[\text{pH}(\psi_d^M)
  - \text{pH}]$ , $\psi_d^M := \psi_d$;
  
  else  $\psi_0^m := \psi_d$ ; $\psi_d := (\psi_d^m +  \psi_d^M)/2$;

  Go to step 2.
\end{enumerate}
If instead of the pH, one sets the bare charge as a constant parameter,
pH and $\text{pH}(\psi_d)$ in step 5 have to be replaced by
$\sigma$ and $\sigma(\psi_d)$, respectively.

\paragraph{The polydisperse cell model} can be advantageously
solved, not by imposing a particle volume fraction, but, instead, by
setting the same electrostatic potential  $\psi_c$ at the cell edge for all colloidal families,
i.e. the condition defined by Eq. \ref{closcm}. The
particle volume fraction is then calculated \textit{a posteriori}. That is, for a given $\psi_c$ the
corresponding set of cell radii $\{R_{c,p}\}_{p=1,\dots,n_p}$ is
calculated iteratively by solving Eq. \ref{PBrad} in such a way as the
condition defined by Eq. \ref{closcm} is respected and by imposing the
boundary conditions defined by
Eqs. \ref{bccm}, \ref{bcch2CM}, \ref{pHregul}, and \ref{stern}. $\Phi$ is then
calculated with  Eq. \ref{vfCM}. The proposed algorithm is summarized below:
\begin{enumerate}
\item Choose a potential at the cell edge $\psi_c$.
\item For each colloidal species $p$ choose a first guess  $R_{c,p}$, within a range [$R_{c,p}^m$, $R_{c,p}^M$] .
\item For each $p$ solve the PBE for a given pH, see~\autoref{PBE}.
\item For each $p$ extract $\psi_p(R)$.
\item If $|\psi_p(R_{c,p})-\psi_c|$ is small enough, go to step 7.
\item Dichotomy step:

  if $\sign[\psi_p(R_{c,p})-\psi_c] = \sign[\psi_p^M(R_{c,p}) -
    \psi_c]$ , $R_{c,p}^M := R_{c,p}$;
  
  else  $R_{c,p}^m := R_{c,p}$ ; $R_{c,p} := (R_{c,p}^m +R_{c,p}^M)/2$.

  Go to step 3.
\item Calculate ${Z^{\ast}}_{p=1,\dots,n_p}$, $\Phi$ (Eq. \ref{vfCM}), and $P$ (Eq. \ref{prjm})
\end{enumerate}

\paragraph{The polydisperse jellium model} is simpler to solve since
it eliminate the cell radius and the corresponding
adjustment. In fact, no iteration is required if it is solved from a
set value of the background charge. Alternatively, a
given $\Phi$ can be achieved by iteratively adjusting the background
charge. The proposed algorithm reads:
\begin{enumerate}
\item Choose a background charge potential $\psi_D$. (see Eq. \ref{defphiD}).
\item For each colloid $p$ compute the potential profile $\psi_p(r)$ at a given pH (see~\autoref{PBE}) for a given $\psi_D$.
\item Calculate ${Z^{\ast}}_{p=1,\dots,n_p}$, then $\Phi$ (Eq. \ref{vfRJM}), and $P$ (Eq. \ref{pcm}).
\item Optionally, if a given $\Phi_{goal}$ is imposed,  inverse
  Eq. \ref{vfRJM} with $\Phi=\Phi_{goal}$ to obtain a new $\psi_D$ and go
  to step 2, unless $|\Phi(\psi_D) - \Phi_{goal}|$ is small enough.
\end{enumerate}

\paragraph{The application to continuous size distribution} of the PCM
and PRJM takes advantage of the continuous variation of the effective charge
with its curvature and is simplified with the proposed algorithm
where the particle volume fraction is not an input parameter but
calculated \textit{a posteriori}. For relatively small adimensional curvatures the charge scales linearly with $1/\kappa
R_p$, $\sigma^{\ast}(R_p)=\sigma^{\ast}_{plane}(1+A(\kappa R_p)^{-1})$, while for large $1/\kappa
R_p$ it scales quadratically,
$\sigma^{\ast}(R_p)=\sigma^{\ast}_{plane}(1+A(\kappa
R_p)^{-1}+B(\kappa R_p)^{-2})$, see the results section for more details.

\paragraph{The source code} for the PRJM and PCM, along
with examples, is available at this address: \url{https://github.com/guibar64/polypbren}.

\subsection{Suspensions and model details}
\label{sec:supensions_details}

As specified earlier, we focus in this paper on polydisperse suspensions
of titratable silica (SiO$_2$) nano-particles with continuous
size distribution. As silicon is one of the major elements of the Earth's
crust, the chemistry and, of particular interest here, the surface chemistry of SiO$_2$ are quite well defined and
documented. The surface of SiO$_2$ is covered with titratable silanol
groups with a surface density $d_s$. These titrate with pH
according to the following equilibrium reaction
\begin{equation}
\label{equilsi}
\ce{Si-OH <=> M-O- + H+} .
\end{equation}
The corresponding equilibrium constant, p$K_a$, as well as the
thickness of the Stern layer, $\lambda_{Stern}$ and d$_s$ were fitted against experimental titration data as
obtained by Dove \textit{et al.} \cite{Dove:05}, see Sect. \ref{sec:charging_silica}. These
parameters were then maintained constant for all other
calculations. 
A large number of the calculations were made with size distributions
corresponding to commercially available silica suspensions, namely
Ludox HS40 and TM50 suspensions, thereafter denominated HS40 and TM50,
respectively. They are described in detail elsewhere\cite{Goehring:17}.
In short, we used a gamma distribution for the HS40 and a normal distribution for the TM50.
In particular, for HS40, an average radius of 8.14 nm and a polydispersity of
14 \% and for TM50, $\langle R_p \rangle=12.1\text{ nm}$ and a polydispersity of 12
\% were used. Calculations were also performed with various
distribution shapes and varying polydispersities as indicated in the text.

All the calculations were performed at a finite concentration (5 mM
for most of them) of a mono-monovalent salt, $T = 300\text{ K}$ and $\lambda_B =$
0.7105~nm.

\section{Results}
\label{results}
Before comparing the generalized cell and renormalized jellium models,
the charging process of silica is presented and modeled to extract the
ionization constant, the density of titratable sites and the thickness
of the Stern layer.

\subsection{Charging process of silica}
\label{sec:charging_silica}

\begin{figure}[!htbp]
  \includegraphics[width=\figurewidth]{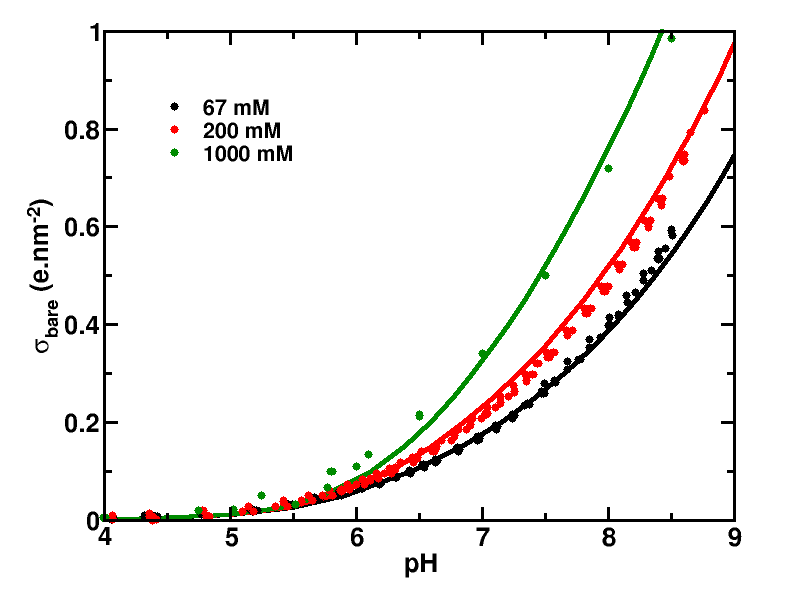}
  \caption{\label{fig1_fitsigvspH} Comparison of experimental and
    simulated bare surface charge density versus
    pH for silica in aqueous solution at different NaCl
    concentrations. The simulations are represented by solid lines,
    the experimental data by points. The salt concentrations are
    67~mM (black), 200~mM (red), 1000~mM (green). The experimental
    data are from Dove \textit{et al.}\cite{Dove:05}.
      } 
\end{figure}

Figure \ref{fig1_fitsigvspH} presents the titration curve of silica in NaCl salt solution
at three different concentrations, these data were obtained by Dove \textit{et al.} \cite{Dove:05}.
The charge density (in absolute values) increases with
increasing pH due to the progressive dissociation of the silanol
groups. $\sigma$ is also seen to increase with the salt concentration as
a result of a greater screening of the electrostatic repulsion between
charged sites.
The following set of parameters, p$K_a$ = 7.7, $d_{site}$ = 5.55
nm$^{-2}$ and $\lambda_{Stern}$ = 0.107 nm, is found to provide
a good description of the charging process of silica. Note that these
parameters were obtained with Eq. \ref{pHregul} assuming a planar
surface in the limit of infinite dilution. They are kept constant in the
rest of the manuscript. The surface charge densities of a
planar silica surface for several pH values and conditions used throughout
this study are given in Table \ref{tabsplane}.

\begin{table}[!htbp]
  \begin{ruledtabular}
  \begin{tabular}{cc}
    pH & Surface charge ($\text{e.nm}^{-2}$)  \tabularnewline
    \hline 7 &  0.0816  \tabularnewline
    8 &  0.18  \tabularnewline
    9  & 0.365  \tabularnewline
    9.5  & 0.508  \tabularnewline
    10  &  0.695\tabularnewline
    10.5  &  0.932\tabularnewline
  \end{tabular}
  \end{ruledtabular}
  \caption{\label{tabsplane} pH and bare surface charge density
    calculated for a planar silica surface in a monovalent salt
    solution with $c_s=5\text{ mM}$, $\lambda_B
    =  0.7105\text{ nm}$, $pKa = 7.7$, $d_{site} = 5.55 \text{
      nm}^{-2}$, $\lambda_{Stern} = 0.107 \text{ nm}$.}
\end{table}

\begin{figure}[!ht]
  \includegraphics[width=\figurewidth]{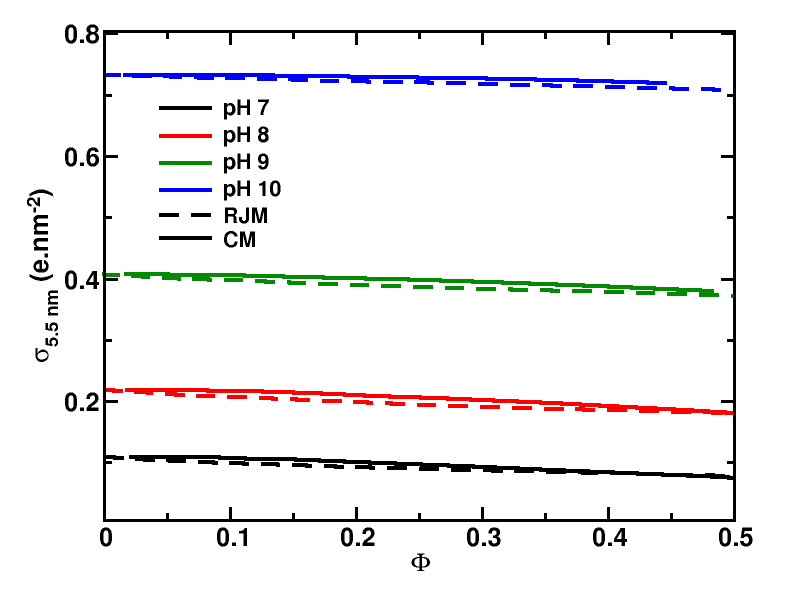}
  \caption{\label{fig2_sbare_R5p5_vs_vf} Calculated bare surface charge density
    for silica particles with $R_p$ = 5.5 nm when varying the particle
    volume fraction of a polydisperse HS40 suspension in monovalent
    salt solution at different pHs: 7 (black), 8 (red), 9 (green), and
    10 (blue). The results are presented both with the polydisperse CM
    (solid lines) and RJM (dashed lines) approximations. The salt
    concentration is set to 5 mM.
  }
\end{figure}

$\sigma$ titrates with pH but also regulates as the
particle volume fraction increases. The drop of $\sigma$ with
$\psi$ is illustrated in Fig. \ref{fig2_sbare_R5p5_vs_vf} for the
particle family of radius 5.5~nm in an HS40 suspension
(polydispersity 14\%) dispersed in a monovalent salt solution. This
trend is similar in the PCM and PRJM and is explained by the co-ion
exclusion which effectively mimics the strong interactions of the
colloidal particles with their neighbors. The calculated
$\sigma$, although close, tends to be larger within the PCM than
the PRJM. This discrepancy increases as the pH is depressed ($<$10\% at
pH 7, and $<$1\% at pH 10).

\begin{figure}[!h]
  \subfigure[\;CM]{\includegraphics[width=\figurewidth]{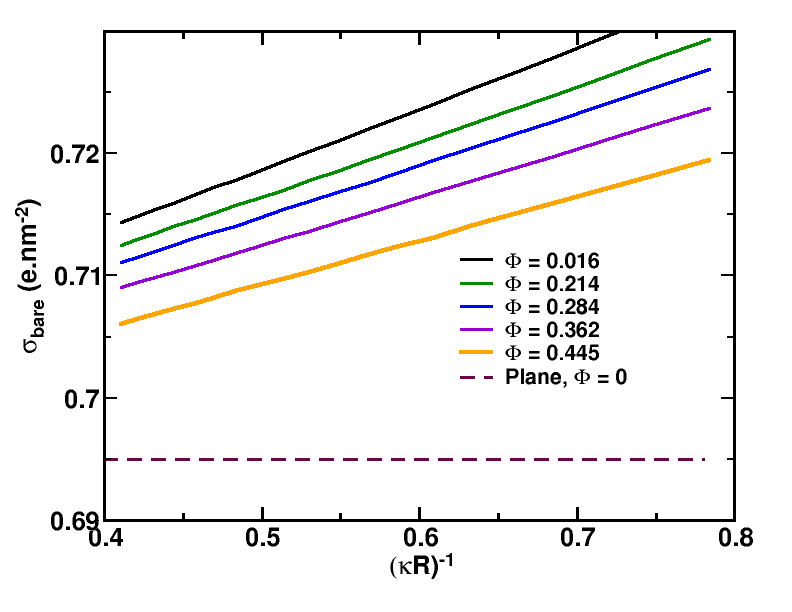}}
  \subfigure[\;RJM]{\includegraphics[width=\figurewidth]{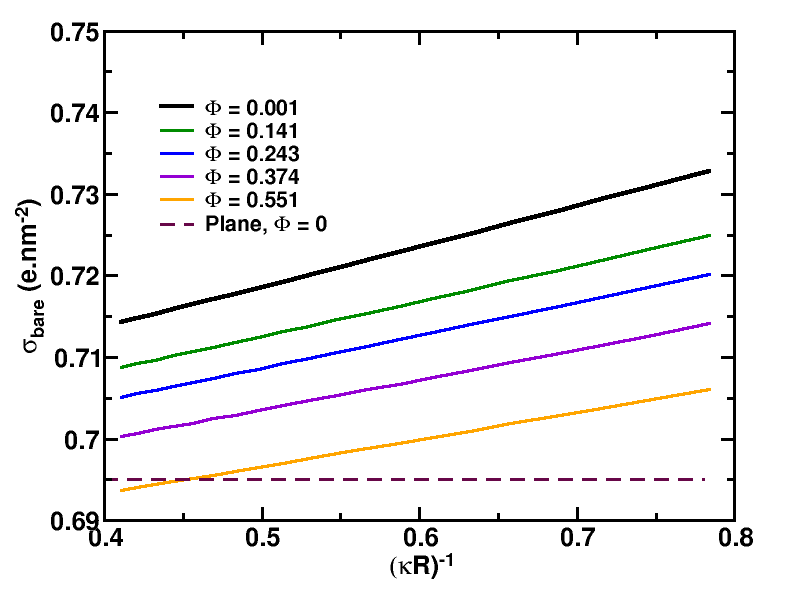}}
  \caption{\label{fig3_sbare_vs_kr_ph10} Bare surface charge
    $\sigma$ versus dimensionless curvature $(\kappa
    R)^{-1}$ at pH 10 and several volume fractions (see legend), for
    the PCM (a) and the PRJM (b). The surface charge for a planar
    surface at infinite dilution ($\Phi=0$) are displayed in both
    cases with a purple dashed line.}
\end{figure}

A result of the charge titration is also the curvature dependence of the
particle charging. In particular, Abbas and coworkers\cite{Abbas:08}
showed that there is a considerable increase in the surface charge
density for particles smaller than 10 nm in diameter. The rise in charge
up with particle curvature can be
understood as an enhanced screening of small sized
particles by counter-ions as compared to that of large particles. This
is illustrated in Fig. \ref{fig3_sbare_vs_kr_ph10} as a function of the
dimensionless curvature $(\kappa R_p)^{-1}$ at pH=10 for various
$\Phi$ of the HS40 silica particle dispersion. The calculations are
performed both in the PCM and PRJM approximations and are compared to
the planar case at infinite dilution. Interestingly, the
curvature dependence of $\sigma$ is found to vary linearly
with $(\kappa R_p)^{-1}$. This can be explained by the Taylor development
of $\sigma$ with respect to $(\kappa R_p)^{-1}$ which in the limit of 
$\kappa R_p \gg 1$ takes the form
\begin{equation}
  \label{barescaling}
  \sigma(R_p)=\sigma_{plane}(1+A_0(\kappa R_p)^{-1}),
\end{equation}
where
  $\sigma_{plane}$ is the charge density of a planar surface and $A_0$ gives
the slope. Note that
here it also applies to relatively small $(\kappa R_p)^{-1}$. It should be
mentioned, however, that in the pH region of large charge regulation,
typically for pH values close to p$K_a$ \cite{Lund:05b}, the linear relationship only holds
for $\kappa R_p > 2$, not shown. The slope of $\sigma(1/\kappa R_p)$ is seen to decrease slightly with the particle volume fraction
as a result of increasing counter-ion screening which tends to moderate,
in relative terms, that due to curvature. Also, in the large
$\Psi$ domain, the $\sigma$ of the small particles becomes lower than $\sigma_{plane}$
in the reference state (i.e. $\Psi=0$), see
e.g. Fig. \ref{fig3_sbare_vs_kr_ph10}-b.
In the infinite dilution limit where a
Grahame relation for the nonlinear
PBE in the spherical geometry has been recently obtained \cite{Carvalho:16}, an analytical expression for
$A_0$ can be found. It reads
\begin{equation}
  A_0 = \frac{ \frac{1}{\cosh^{2}(\psi_{0,plane}/4)} + 
  \frac{C \sigma_{plane} \kappa \lambda_{Stern}}{2 q_s \tanh (\psi_{0,plane}/2)} }
  {1 + \frac{1/(1-\alpha_{plane}) + C \sigma_{plane}}{2 q_s \tanh (\psi_{0,plane}/2)} }.
\end{equation}
A detailed development is given in the SI. 

\begin{figure}[!ht]
  \subfigure[\;]{\includegraphics[width=\figurewidth]{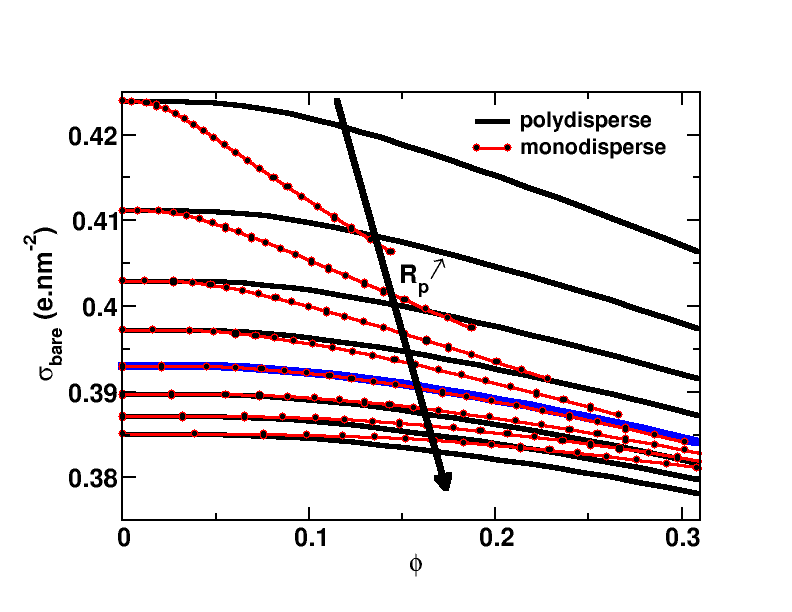}}
  \subfigure[\;]{\includegraphics[width=\figurewidth]{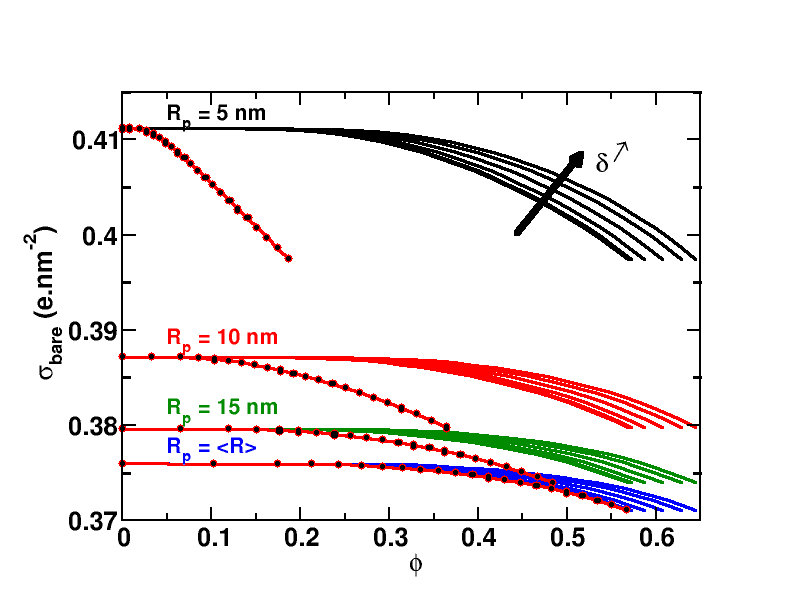}}
  \caption{\label{fig4_sbare_vs_polydisp_ph9} Bare surface charge density,
    $\sigma$, for particle suspensions immersed in a 5~mM 1-1
    salt solution at pH 9 for various particle sizes, size
    distributions and particle volume fractions a) $\sigma$ for
    particles of various $R_p$ (4, 5, 6, 7, 8, 9, 10 and 11 nm) within
    HS40 suspensions in comparison with the corresponding monodisperse
    cases. The blue line gives that of particles with $R_p = \langle
    R_p \rangle$. b)
    The same as in (a) but for particle suspensions having a normal radii
    distribution with $\langle R_p \rangle$ = 20 nm at different polydispersities
    $\delta$ (5, 10, 20, 30, 40 and 50 \%). $\sigma$ of
    monodisperse particle suspensions (red lines with symbols) are
    also given for comparison.}
  \end{figure}

The influence of polydispersity on the bare surface charge density
of different particle families, i.e. with different $R_p$, is illustrated in
Fig. \ref{fig4_sbare_vs_polydisp_ph9} which compares the case of
polydisperse and monodisperse suspensions for various particle size
distributions. Interestingly, for particles with $R_p$ equal to the
mean value of the distribution, $R_p=\langle R_p \rangle$, the
polydispersity, when it is relatively small ($<$15\%), has virtually no impact
on $\sigma$. As could be expected, this is the same for
infinitely diluted suspensions whatever the particle family or
polydispersity, see Fig. \ref{fig4_sbare_vs_polydisp_ph9}-b. On the
contrary, as $R_p$ departs from $\langle R_p \rangle$, the $\sigma$ of mono- and
poly-disperse suspensions can clearly be seen to differentiate and this
differentiation steps up with $\Phi$ and the departure from $\langle R_p \rangle$. The
polydispersity effect is more pronounced for the small particles of
the size distribution. In addition, polydispersity yields them higher charges (compared
to monodispersity) which monotonically increase with it. The opposite is found for the large particles.

\subsection{Renormalized parameters}
\label{sec:ren_params}

\begin{figure}[!h]
  \subfigure[\;\label{fig5_keffokvsvf}]{\includegraphics[width=\figurewidth]{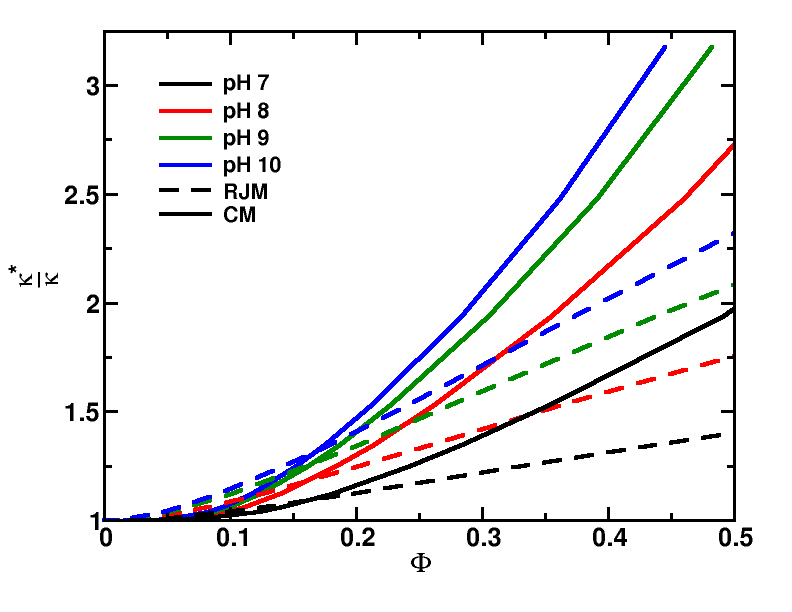}}
  \subfigure[\;\label{fig5_R5p5_vs_vf}]{\includegraphics[width=\figurewidth]{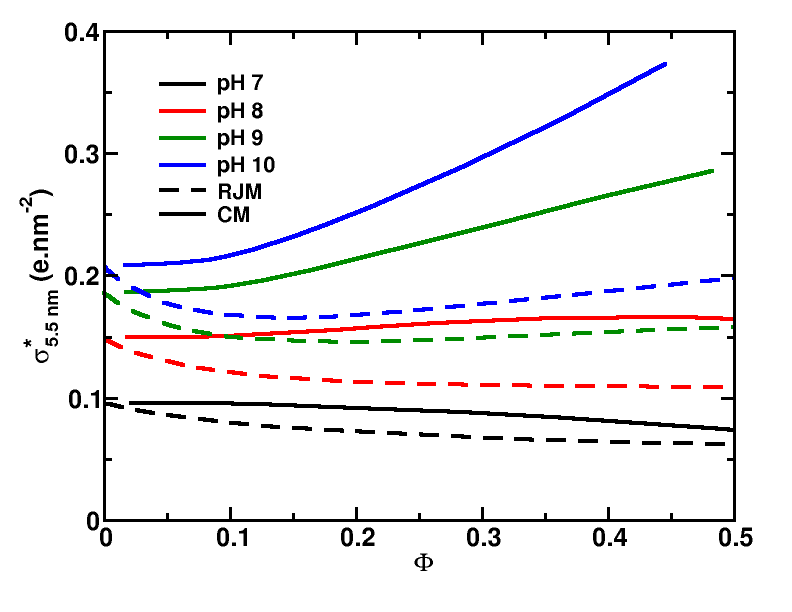}}
  \caption{\label{fig5_keffsbare} a) Relative effective inverse screening
    length  $\kappa^{\ast}/\kappa$, and b) effective surface
    charge density $\sigma^{\ast}$ versus volume fraction $\Phi$, for
    particles with $R_p=$5.5~nm of the HS40 dispersion, at the
    following pHs: 7 (black), 8 (red), 9 (green) and 10 (blue). The
    ionic strength is set to 5 mM. The
    full lines give the results of the PCM while the dashed lines those of the PRJM.
  } 
\end{figure}

So far we have seen that the bare surface charge densities as obtained
from the PCM and PRJM approximations are very similar whatever the
particle size distribution or particle volume fraction. This is no
longer true for the renormalized charge and screening length as shown in
Fig.\ref{fig5_keffsbare}. These parameters are calculated for HS40 suspensions at various $\Phi$.

$\sigma^{\ast}$ obtained within the PRJM is found
to be lower than that within the PCM, whatever the $\Phi$ and all the more so as
pH increases, that is as the effective charge approaches saturation. The same is observed for $\kappa^{\ast}$
but in the domain of large $\Phi$ ($\Phi\gtrsim 0.15$) while
the opposite is found, that is $\kappa^{\ast}_{RJM} >
\kappa^{\ast}_{CM}$, in the dilute and semi-dilute regimes ($\Phi \lesssim0.15$).
These results are consistent with those obtained by
Trizac \textit{et al.} with monodisperse suspensions, see
Refs \onlinecite{Trizac:04, Dobnikar:06}, but are here exacerbated by the
polydispersity. In particular, $\kappa^{\ast}$ values of both models are
found not to converge in the limit of large $\Phi$, the domain of
counter-ion dominated systems (supposedly equivalent to the salt free
case), but, instead, to become increasingly divergent even at low pH
values. Note that in the salt free case (not shown), $\sigma^{\ast}$ is
still distinctly lower in the RJM, but the $\kappa^{\ast}$ of both models
are found to be similar in the domain of low $\Phi$ ($<0.2$).

\begin{figure}[!h]
  \subfigure[\;\label{fig6_seffplanevsvf}]{\includegraphics[width=\figurewidth]{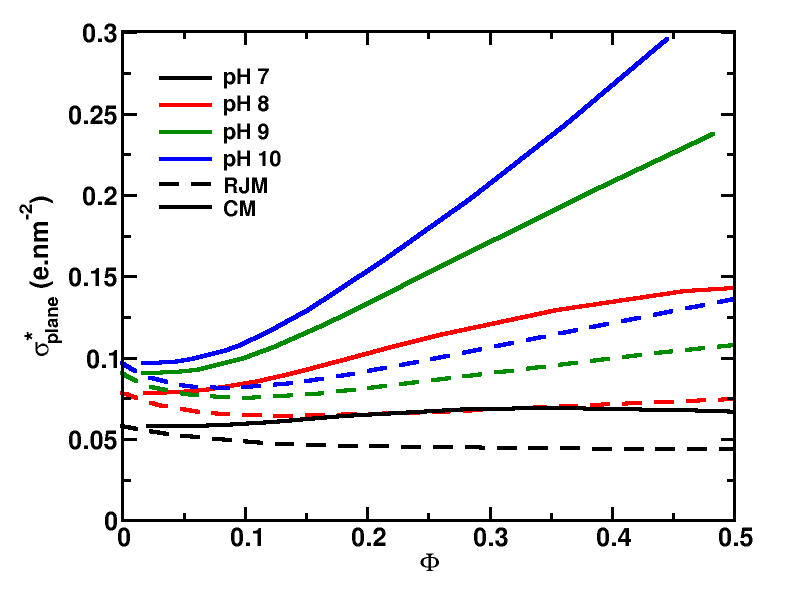}}
  \subfigure[\;\label{fig6_Avsvf}]{\includegraphics[width=\figurewidth]{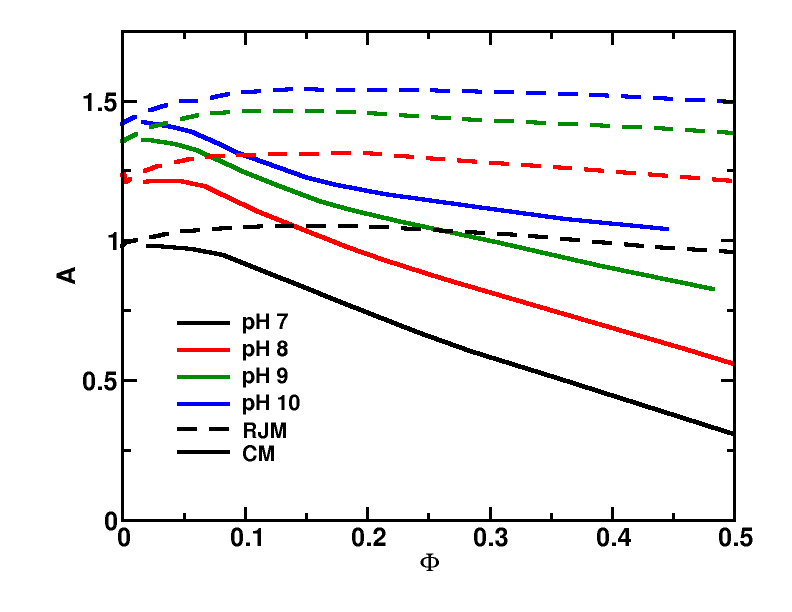}}
  \caption{\label{fig6_seffplaneA} Effective charge density of the planar surface $\sigma^{\ast}_{plane}$ 
and factor $A$ versus volume fraction $\Phi$. Those were worked out for the HS40 distribution, and
for the following pHs: 7 (black), 8 (red), 9 (green), and 10 (blue), using the PCM (full lines), and the PRJM (dashed lines)
.}
\end{figure}

In the same way as for the bare charge density, the effective surface charge
density can be accurately approximated by means of an affine
function of $(\kappa^{\ast}R_p)^{-1}$, that is,
\begin{equation}
\label{decompseff}
\sigma^{\ast} = \sigma^{\ast}_{plane} \left ( 1 + A (\kappa^{\ast}R_p)^{-1}  \right ),
\end{equation}
where $\sigma^{\ast}_{plane}$ is the effective surface charge density
of the confined planar surface in the same conditions (pH and $\Phi$,
i.e. same edge potential for the PCM and same background charge for the
PRJM) and $A$ is a dimensionless coefficient which measures the impact of the size of the particle.
 $\sigma^{\ast}_{plane}$ and $A$ depend on the model, pH, particle size distribution and
volume fraction. Equivalently, Eq. \ref{decompseff} can be written in
terms of effective charge, that is $Z^{\ast}/R_p=
4\pi\sigma^{\ast}_{plane}/\kappa^{\ast}(\kappa^{\ast} R_p+A)$.
In the case of no added salt, after noting that
$\sigma^{\ast}=\frac{\gamma\kappa^{\ast}}{\pi \lambda_B}$, where $\gamma$ is a
coefficient which varies with $\sigma$ and $\Phi$ \cite{Bocquet:02}, it was found that
$Z^{\ast}$ scales linearly with the ratio $R_p/\lambda_B$. Such linear scaling was
verified experimentally for deionized colloidal suspensions in the infinite dilution
limit, in the semi-dilute regime and in the concentrated regime by measurements of electrophoretic mobility of isolated
colloids\cite{Garbow:04,Strubbe:06}, conductivity of colloidal liquids and elasticity of
colloidal crystals\cite{Wette:02}, respectively. 

In the case of added salt ($\kappa R_p \gg 1$) and infinite dilution limit,
where an analytical expression of the electrostatic potential solution
of the non linear Poisson Boltzmann theory has been obtained
\cite{Shkel:00, Aubouy:03}, an analytical approximation of the
coefficient A for non titrating colloids
can be obtained and reads,
\begin{equation}
A=\frac{1}{2}\left( 5 - \frac{\gamma^4+3}{\gamma^2+1}\right),
\end{equation}
where $\gamma=\sqrt{1+x^2}-x$ and $x=\frac{\kappa}{2\pi \lambda_B\sigma}$.
The approximation is asymptotically exact in the limit of
large R, see the SI for a detailed development. Finally, since
$\gamma$ goes to 1 when $\sigma \rightarrow \infty$ one finds
$A_{sat}=3/2$ at the saturation of the colloidal charge.

Fig. \ref{fig6_seffplaneA} shows the PCM and PRJM results of the coefficient A and
the effective charge density of the plane, $\sigma^{\ast}_{plane}$, versus $\Phi$ on HS40 at different
pH and a set ionic strength of 5 mM. Not surprisingly, $\sigma^{\ast}_{plane}$ follows the same
trend as for $\sigma^{\ast}_{5.5\text{ nm}}$,
c.f. Fig. \ref{fig5_R5p5_vs_vf}. In particular, for pH values greater
than the p$K_a$ (pH $> 7.7$) PRJM's $\sigma^{\ast}_{plane}$ systematically
shows a non monotonous behavior with respect to $\Phi$ with a minimum
around $\Phi\approx 0.1$. Within the PCM, on the other hand,
$\sigma^{\ast}_{plane}$ continuously rises with $\Phi$. The difference in
behavior in $\sigma^{\ast}_{plane}$ between the two models is
reminiscent of that of $\sigma^{\ast}_{plane}$ at saturation which
follows the same qualitative trend, see
e.g. Ref. \onlinecite{Pianegonda:07}. Indeed, in these conditions of pH,
$\sigma_{plane}$ is generally larger than $\sigma^{\ast}_{plane,sat}$.
For pH values lower than the p$K_a$ and at
relatively high $\Phi$ the PCM's
$\sigma^{\ast}_{plane}$ also shows a drop due to the regulation of
the bare charge density which becomes much smaller than $\sigma^{\ast}_{plane}$ at saturation.

This qualitative difference is echoed in the coefficient $A$ which shows a maximum value in the
PRJM and not in the PCM, see Fig. \ref{fig6_seffplaneA}-b. Indeed, $A$
 varies between a maximum value $A_{sat}$, at the saturation of the
 charge and a minimum value $A_{0}$ when $\sigma^{\ast}=\sigma$,
 c.f. Eq. \ref{barescaling}. It naturally follows that
 $\sigma^{\ast}(R_p)$ desaturates as $\Phi$ further increases and approaches
 the ideal planar limit where the effective charge is proportional to
$R_p^2$. In this respect, the PCM's $A$ values decrease faster with $\Phi$
than is the case in the PRJM. In addition, for non titrating
surfaces $A_0=0$ ($\sigma(R_p)=\sigma_{plane}$) and, as we
  have seen above, $A_{sat}~1.5$ at $\Phi=0$ both for titrating and non titrating surfaces.

\begin{figure}[!htbp]
  \subfigure[\;CM]{\includegraphics[width=\figurewidth]{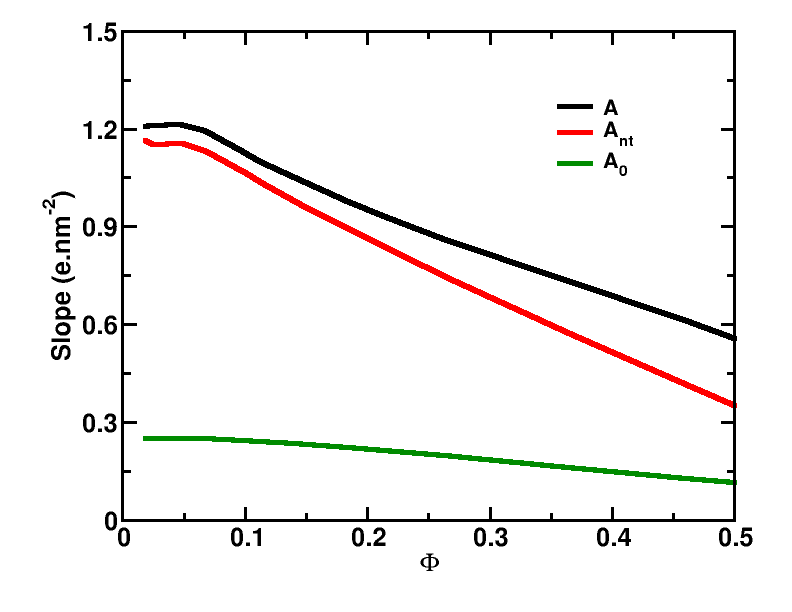}}
  \subfigure[\;RJM]{\includegraphics[width=\figurewidth]{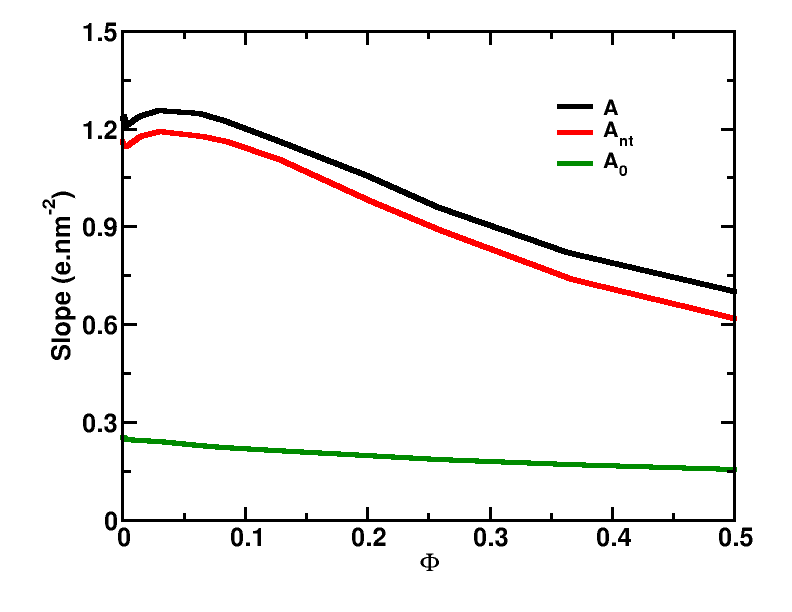}}
  \caption{\label{pente_pH8} Slopes of the linear variation of
    $\sigma^{\ast}$ with $(\kappa R_p)^{-1}$ of a titrating, $A$, and a
    non-titrating colloidal particle, $A_{nt}$. That of the bare
    surface charge density, $A_0$, is also given as a reference. The
    slopes are given for both the PCM (a) and PRJM (b). Calculations are
    performed for silica HS40 dispersions in equilibrium with a bulk
    solution containing 5mM of 1-1 salt and at pH 8. In the non
    titrating case, the particles are given a surface charge density
    equal to that of the planar silica surface in the same conditions.}
\end{figure}

This points to the fact that for non titrating colloids, with a $\sigma(R_p,\Phi)$
equal to that of a titrating planar surface in the
same conditions ($\sigma_{plane}(\Phi,\mathrm{pH})$), the corresponding
coefficient, $A_{nt}$, is lower than $A$ for charge regulating particles. In other words, $A$ is a
function of $A_0$ and $A_{nt}$, see Figure \ref{pente_pH8}. In the limit of small variations of
$\sigma$, one can further give an analytic approximation for the
dependence of $A$ on $A_0$ and $A_{nt}$ which reads,
\begin{equation}
A\approx
A_{nt}+\frac{\kappa^{\ast}}{\kappa}A_0\frac{\sigma_b}{\sigma^{\ast}}\frac{\mathrm
  d \sigma^{\ast}}{\mathrm d \sigma_b} .
\end{equation}
Close to the saturation of the effective charge as well as in the limit of small $\sigma_p$ the last expression reduces to
 $A\approx A_{nt}+\kappa^{\ast}/\kappa A_0$. 
 
Not shown here is how the ionic strength affects both
$\sigma^{\ast}_{plane}$ and $A$. This is already well documented in the literature
in the case of monodisperse suspensions, see
e.g. Refs. \onlinecite{Belloni:98, Groot:91, Stevens:96}. Not surprisingly, the same qualitative behavior is
found in polydisperse suspensions. That is, $A$ drops and
$\sigma^{\ast}$ rises when ionic strength increases. It is also
is easy to infer from Eqs. 8-22 and Fig. \ref{fig4_sbare_vs_polydisp_ph9} that an increase in the
polydispersity gives rise to a shift in $\sigma^{\ast}_{plane}$ and $A$
values to larger $\Phi$ values.

In conclusion of this section, we have seen that the well-known linear
scaling of the effective charge with
$(\kappa R_p)^{-1}$ is also verified in the case of polydisperse and charge
regulating colloids for all $\Phi$. In practice, this means that a complete force
field for these suspensions can be obtained at relatively low
computational cost. Indeed, this amounts to calculating $A$, $\sigma^{\ast}_{plane}$
and $\kappa^{\ast}$ with a few $R_p$ values
(in principle two are enough) at set values of $\psi_c$ (in the PCM)
or $Z_{back}$ (in the PRJM), and post-calculating $\Phi$ given
the particle size distribution (continuous or not).

\subsection{Osmotic pressure}
\label{sec:osmotic}

\begin{figure}[!ht]
    \includegraphics[width=\figurewidth]{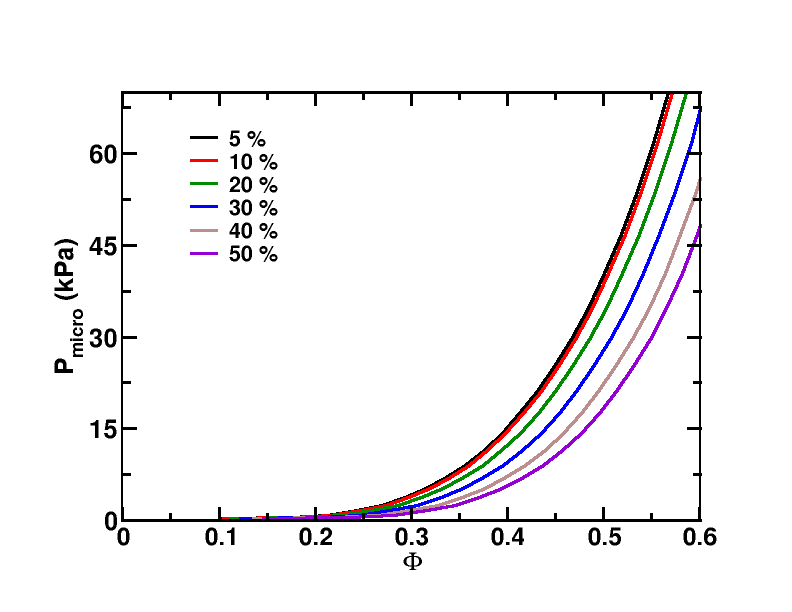}
  \caption{\label{Pmicronormal} Simulated microion contribution to the osmotic
    pressure of titratating silica particle dispersions as a function
    of the particle volume fraction for varying polydispersities. The
    calculations are performed within the PCM approximation. The
    silica particles are suspended in a 5~mM 1-1 salt solution at pH
    9. The particles present a normal size distribution with $\langle R_p \rangle =$
    20 nm. The polydispersity is changed as indicated in the
    legend.
  }
\end{figure}

In this section the effect of the polydispersity on osmotic
pressure is discussed. Finally, the validity of the PRJM and PCM will be
discussed in light of experimental equation of states for various
commercial silica dispersions.

Figure \ref{Pmicronormal} gives the microion contribution to the total osmotic
pressure, $P_{micro}$ as obtained from the PCM with different
polydisperse suspensions having a normal size distribution of the same
mean particle size $\langle R_p \rangle =$ 20
nm but of varying polydispersities. $P_{micro}$ is found to decrease
as $\delta$ increases. The drop in $P_{micro}$ is significant above
10\% of polydispersity. The PRJM exhibits the same qualitative behavior
(not shown).

\begin{figure}[!h]
  \subfigure[\;]{\includegraphics[width=\figurewidth]{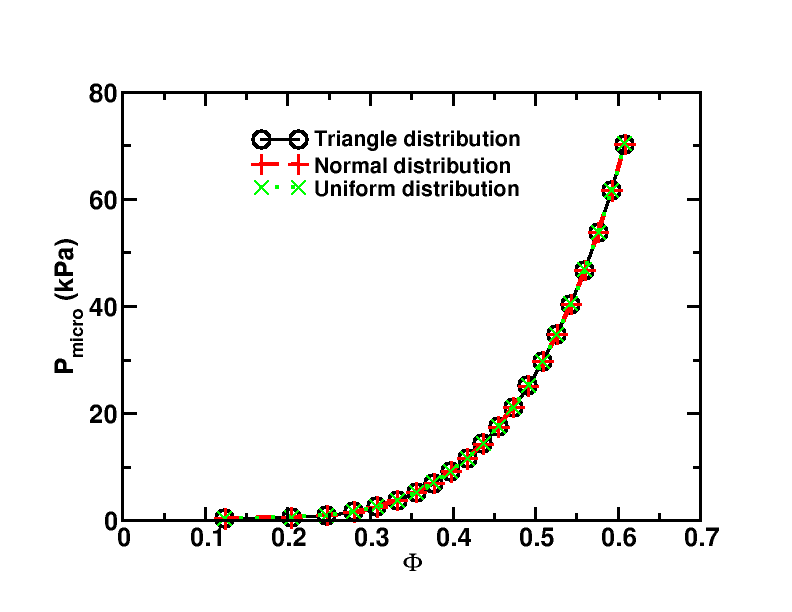}}
  \subfigure[\;]{\includegraphics[width=\figurewidth]{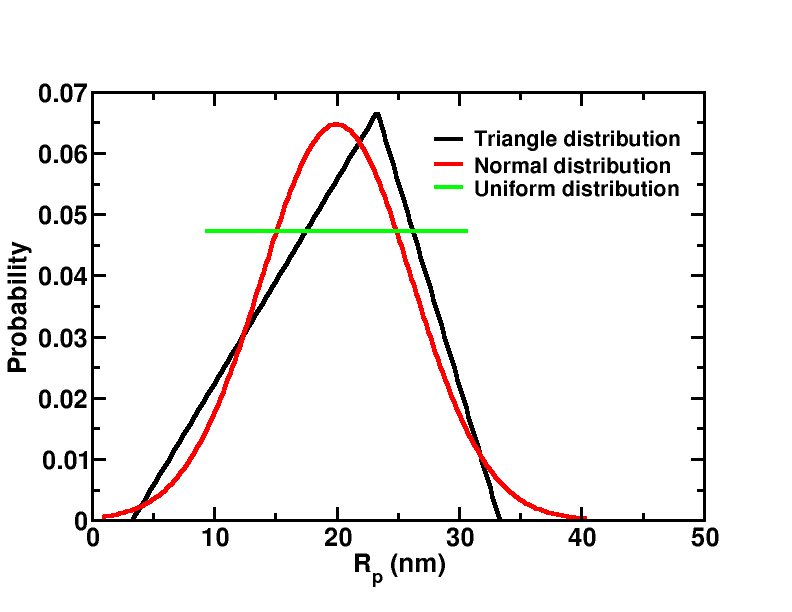}}
  \caption{\label{Pmicrovar} (a) CM calculations of the micro-ions
    osmotic pressure, $P_{micro}$, for silica dispersions with varying
    shapes of particle size distribution but with identical $\langle R_p \rangle$
    (20 nm) and $\langle R_p^3 \rangle$ (21.77$^3$ nm$^3$). The
    silica particles are suspended in a 5~mM 1-1 salt solution at pH
    9. Note that for the normal distribution $\delta=31\%$. The
    distribution is changed as indicated in the legend. (b) The
    triangular, normal and uniform size distributions used.
  }
\end{figure}

The shape of the particle radius distribution is further found to have only a minor effect on
$P_{micro}$. This is all the more true as the particle size
distributions are chosen so as to have identical $\langle R_p^3 \rangle$ and
$\langle R_p \rangle$. As shown in Figure \ref{Pmicrovar}-a for three different distribution
shapes, when these conditions are met
the $P_{micro}$ thus obtained can hardly be distinguished. This behavior
is a direct consequence of the geometrical definition of the particle
volume fraction, see Eq. \ref{phi} combined with the very slow
variation in the water layer thickness,
$R_{cell}-R_p$, with the particle radius. In the limit of large $\kappa R_p$, $R_{cell}-R_p$ becomes constant.

\begin{figure}[!h]
  \subfigure[\;]{\includegraphics[width=\figurewidth]{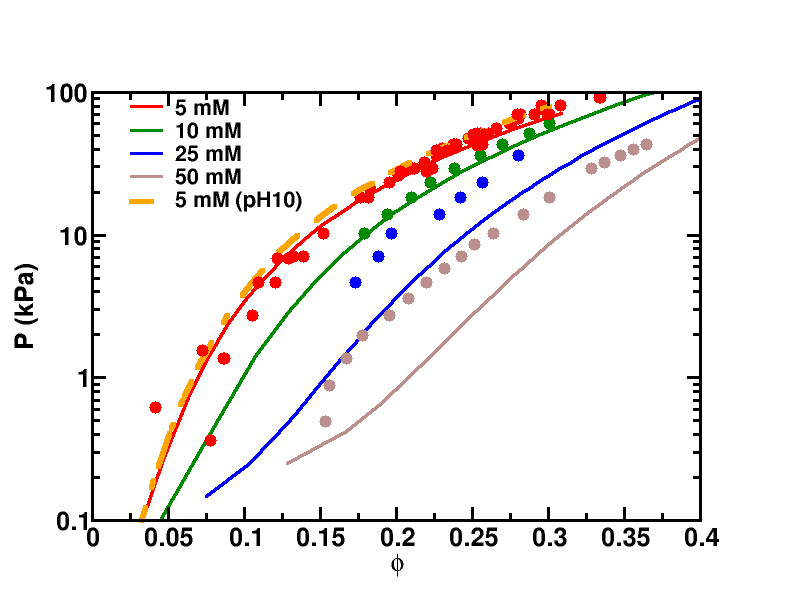}}
  \subfigure[\;]{\includegraphics[width=\figurewidth]{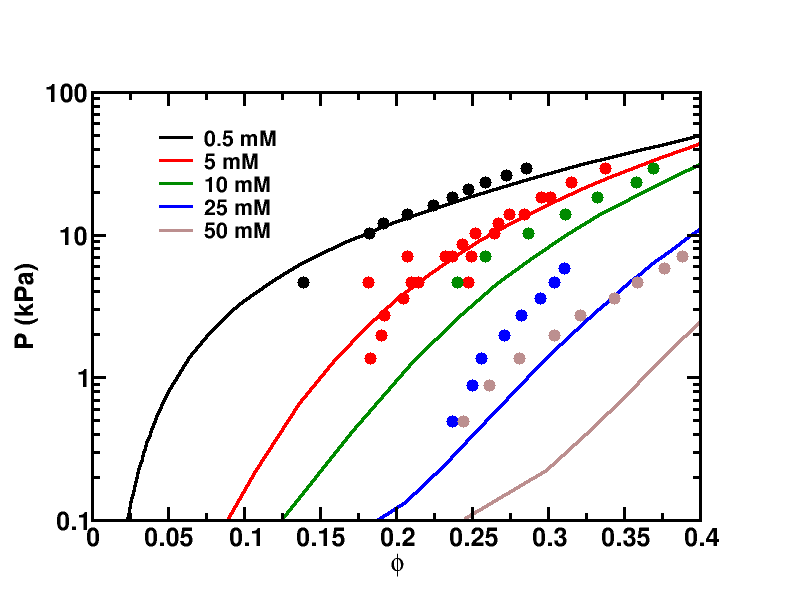}}
  \caption{\label{fig8_eos} Experimental equation of
    state for the (a) HS40 and (b) TM50 silica dispersions in
    comparison with the micro-ion pressure calculated by the polydisperse cell
    model at various bulk concentrations of monovalent salt and pH
    9. The results for the polydisperse RJM are given in the SI.}
\end{figure}

Fig. \ref{fig8_eos} compares the experimental equations of state of the
HS40 and TM50 silica dispersions\cite{Goehring:17} with the micro-ion pressure
calculated with the polydisperse cell model at various bulk
concentrations of monovalent salt and pH 9. The osmotic pressure is
seen to increase when the ionic strength of the bulk and the mean
particle radius ($\langle R_p(\text{HS40}) \rangle < \langle
R_p(\text{TM50}) \rangle$) decreases, in good agreement with the
polydisperse cell model. What is more, the PCM is found to give a good description
of the osmotic pressure of the silica dispersions only for the
lowest bulk salt concentrations studied, up to 10~mM for the HS40 and
to 5~mM for
the TM50. This should not come as a surprise since the PCM is known to
neglect the entropic and contact contributions to the osmotic
pressure\cite{Hallez:14,Dobnikar:06}. As discussed by Hallez \textit{et al.}\cite{Hallez:14}
, it is found that the lower the mean
particle size, the larger the validity range of the PCM. The PRJM, on
the other hand, is found to give a poor description of the
experimental osmotic pressure, see SI. Generally, it overestimates the
osmotic pressure at low volume fractions and underestimates them in the
concentrated regime.

\section{Conclusion}
\label{sec:conclusion}

In this paper, we proposed a cell and a
renormalized jellium model to study the thermodynamic properties and
estimate the renormalized parameters to be used in a one-component
model, i.e. $Z^{\ast}$ and $\kappa^{\ast}$, for
polydisperse suspensions of titratable spherical colloids with a continuous size
distribution. We further proposed a simple algorithm and a Nim implementation to solve them. The models are largely inspired by the work of
Torres\cite{Torres:08} on binary mixtures of colloids with
constant charge. PCM and PRJM include a charge regulation, instead of a
constant charge boundary condition, modeled as a simple 1-p$K$-Stern
model. The application of the models to continuous size distributions
was made simple and easy by the linear scaling of both the bare and
effective charges with the adimensional curvature of the particles,
$(\kappa R_p)^{-1}$. For very small $(\kappa R_p)^{-1}$, $\sigma$ and
$\sigma^{\ast}$ scale quadratically. We presented a detailed example of such an analysis in the case of
aqueous suspensions of silica nanoparticles of various size
distributions. Besides being simple, the 1-p$K$ model was found to give a
very good description of the charging behavior for bare silica
surfaces experimentally observed in diluted conditions, in
accordance with previous studies, see e.g. Ref. \onlinecite{Kobayashi:05}. This allowed us to
constrain the surface chemistry parameters of the PCM,
leaving us with two commonly characterized parameters, that is the pH
and the particle size distribution. Both models give the same
qualitative results. Yet, the cell model thus generalized  is found to predict much
more accurately the equations of state of aqueous silica dispersions at
finite salt concentrations. In general, the bare surface density
is found to drop as the density and the radius of the
silica particles increases, due to the charge regulation. In a polydisperse
suspension, the particles of radius $R_p < \langle R_p \rangle$ are further
found to bear a surface charge density significantly greater than that of the same particles at the same
density but in a monodisperse suspension (the opposite occurs when
$R_p > \langle R_p \rangle$). This is all the more true as
polydispersity rises and $R_p$ is small compared to $\langle R_p \rangle$
($R_p >> \langle R_p\rangle$). In other words, the bare charge polydispersity is found
to increase with the size polydispersity. Not surprisingly, the same
trend is found for the effective charge polydispersity. It
should be stressed, however, that a polydispersity of effective
charges is also present in the case of polydisperse particles having the same bare
surface charge density, although less pronounced. Despite these
differences a significant impact on the microion osmotic pressure is
only seen in suspensions of silica particles with very large
polydispersities ($> 15 \%$). One may expect, however, to observe more
clear effects in the micro-structure of these suspensions, even for relatively small polydispersities.

\begin{acknowledgments}
The authors would like to extend their thanks to Lucas Goehring and Joaquim Li for
providing us with their experimental osmotic pressure data and to Bernard
Cabane and Robert Botet for helpful discussions.
\end{acknowledgments}

\bibliography{biblio}

\end{document}